# Object Database Benchmarks

Jérôme Darmont
ERIC, Université Lumière Lyon 2, France

## INTRODUCTION

The need for performance measurement tools appeared soon after the emergence of the first Object-Oriented Database Management Systems (OODBMSs), and proved important for both designers and users (Atkinson & Maier, 1990). Performance evaluation is useful to designers to determine elements of architecture and more generally to validate or refute hypotheses regarding the actual behavior of an OODBMS. Thus, performance evaluation is an essential component in the development process of well-designed and efficient systems. Users may also employ performance evaluation, either to compare the efficiency of different technologies before selecting an OODBMS or to tune a system.

Performance evaluation by experimentation on a real system is generally referred to as benchmarking. It consists in performing a series of tests on a given OODBMS to estimate its performance in a given setting. Benchmarks are generally used to compare the global performance of OODBMSs, but they can also be exploited to illustrate the advantages of one system or another in a given situation, or to determine an optimal hardware configuration. Typically, a benchmark is constituted of two main elements: a workload model constituted of a database and a set of read and write operations to apply on this database, and a set of performance metrics.

## BACKGROUND

### OBJECT DATABASE BENCHMARKS EVOLUTION

In the sphere of relational DBMSs, the Transaction Performance Processing Council (TPC) issues standard benchmarks, verifies their correct application and regularly publishes performance tests results. In contrast, there is no standard benchmark for OODBMSs, even if the more popular of them: OO1, HyperModel, and OO7, can be considered as *de facto* standards.

OO1, also referred to as the "Cattell Benchmark" (Cattell, 1991), was developed early in the nineties when there was no appropriate benchmark for engineering applications such as Computer Aided Design (CAD), Manufacturing (CAM), or Software Engineering (SE). OO1 is a simple benchmark that is very easy to implement. A major drawback of this tool is that its workload model is too elementary to measure the elaborate traversals that are common in many types of object-oriented applications.

The HyperModel Benchmark (Anderson *et al.*, 1990), also referred to as the Tektronix Benchmark, possesses a richer workload model than OO1. This renders it potentially more effective than OO1 in measuring the performance of engineering databases. However, this added complexity also makes HyperModel harder to implement.

OO7 (Carey, Dewitt & Naughton, 1993) reuses the structures of OO1 and HyperModel to propose a more complete benchmark and to simulate various transactions running on a diversified database. It has also been designed to be more generic than its predecessors and to correct some of their known weaknesses. However, OO7 is even harder to implement than HyperModel.

OO1, HyperModel, and OO7, though aimed at engineering applications, are often viewed as general-purpose benchmarks. However, they feature relatively simple databases and are not well suited for other types of applications such as financial, telecommunication, and multimedia applications (Tiwary, Narasayya & Levy, 1995). Hence, many benchmarks were developed to study particular domains, such as client-server architectures (Schreiber, 1994), object clustering (Bancilhon, Delobel & Kanellakis, 1992; Gerlhof *et al.* 1996; Darmont, Petit & Schneider 1998), object-relational systems (Carey, Dewitt & Naughton, 1993; Lee, Kim & Kim 2000), active databases (Zimmermann & Buchmann, 1995), workflow management (Bonner, Shrufi & Rozen, 1995), CAD applications (Kempe *et al.*, 1995), or the study of views in an object-oriented context (Kuno & Rundensteiner, 1995). A fair number of these benchmarks are more or less based on OO1, HyperModel, or OO7.

An alternative to very specific benchmarks resides in generic and tunable benchmarks such as OCB (Darmont & Schneider, 2000). The flexibility and scalability of OCB is achieved through an extensive set of parameters that helps OCB simulate the behavior of the *de facto* standards in object-oriented benchmarking. Furthermore, OCB's generic model can be implemented within an object-relational system easily and most of its operations are relevant for such a system. Hence, it can also be applied in an object-relational context with few adaptations.

Finally, OCB has been recently extended to become the Dynamic Object Evaluation Framework (DOEF), which introduces a dynamic component in the workload (He & Darmont, 2003). Changes in access patterns indeed play an important role in determining the efficiency of a system or of key performance optimization techniques such as dynamic clustering, prefetching, and buffering. However, all previous benchmarks produced static access patterns in which objects were always accessed in the same order repeatedly. In contrast, DOEF simulates access pattern changes using configurable styles of change.

**ISSUES AND TRADEOFFS IN BENCHMARKING**

Gray (1993) defines four primary criteria to specify a good benchmark: (1) *relevance,* it must concern aspects of performance that appeal to the largest number of potential users; (2) *portability,* it must be reusable to test the performances of different OODBMSs; (3) *simplicity,* it must be feasible and must not require too many resources; and (4) *scalability,* it must be able to be adapted to small or large computer systems or new architectures. Table 1 summarizes the characteristics of the main existing benchmarks according to Gray's criteria. It is important to note that these four criteria are in mutual conflict. For instance, the size and complexity of a relevant workload may come in conflict with its feasibility and possibly with portability requirements. Hence, it is necessary to find the right compromise regarding given needs.

Table 1: Comparison of existing benchmarks with Gray's criteria

|  | Relevance | Portability | Simplicity | Scalability |
|---|---|---|---|---|
| OO1 | – – | + + | + + | – |
| HyperModel | + | + | – | – – |
| OO7 | + + | + | – – | – |
| OCB | + + | + | – | + + |

*Strong point: +    Very strong point: ++    Weak point: –    Very weak point: – –*

The *de facto* standards in OODBMS benchmarking all aim at being generic. However, they all incorporate database schemas that are inspired by structures used in engineering software, which finally tailors them to the study of these particular systems. Adapting these benchmarks to another domain requires some work and a derived benchmark that takes into account specific elements often needs to be designed. Hence, their relevance decreases when they are applied in other domains but engineering. A solution to this problem is to select a generic benchmark that can be tailored to meet specific needs. However, there is a price to pay. Genericity is achieved with the help of numerous parameters that are not always easy to set up. Thus, the effort in designing a specific benchmark must be compared to the parameterization complexity of a generic benchmark.

There is also another, very different, "qualitative" aspect of benchmarking that has never really been considered in published benchmarks and that is important for a user to consider when selecting an OODBMS. Atkinson *et al.* (1992), Banerjee and Gardner (1995), and Kempe *et al.* (1995) all insist on the fact that system functionality is at least as important as raw performances. Hence, criteria concerning these functionalities should be worked out.

Finally, there is an issue that is not a scientific one. Carey, Dewitt and Naughton (1993) and Carey *et al.* (1994) pointed out serious legal difficulties in their benchmarking effort. Indeed, OODBMS vendors are sometimes reluctant to see benchmark results published. However, designing relevant benchmarks remains an important task and should still be carried out to help researchers, software designers or users evaluate the adequacy of any prototype, system or implementation technique in a particular environment.

# CONCLUSION AND FUTURE TRENDS

The development of new object database benchmarks is now scarce, mainly because the first generation of OODBMSs failed to achieve any broad commercial success. This failure is largely due to the never-ending issue of poor performance compared to relational DBMSs, which are well optimized and efficient. However, with the development of object-oriented programming both off-line and on-line, the need for persistent objects remains. Object-relational systems are now used more and more frequently (to store XML documents, for instance). Thus, the experience that has been accumulated when designing object-oriented database benchmarks could be reused in this context. The challenge for object store designers is now to produce efficient systems, and sound benchmarks could help them achieve this goal.

# R E F E R E N C E S

Anderson, T.L., Berre, A.J., Mallison, M., Porter, H.H. and Scheider, B. (1990). The HyperModel Benchmark. *International Conference on Extending Database Technology*, Venice, Italy, 317-331.

Atkinson, M.P. and Maier, D. (1990). Perspectives on persistent object systems. *4$^{th}$ International Workshop on Persistent Object Systems*, Martha's Vineyard, USA, 425-426.

Atkinson, M.P., Birnie, A., Jackson, N. and Philbrow, P.C. (1992). Measuring Persistent Object Systems. *5$^{th}$ International Workshop on Persistent Object Systems*, San Miniato (Pisa), Italy, 63-85.

Bancilhon, F., Delobel, C. and Kanellakis, P. (eds.). (1992). "Building an Object-Oriented Database System: The Story of O$_2$". Morgan Kaufmann.

Banerjee, S. and Gardner, C. (1995). Towards An Improved Evaluation Metric For Object Database Management Systems. *OOPSLA 95 Workshop on Object Database Behavior, Benchmarks and Performance*, Austin, USA.

Bonner, A.J., Shrufi, A. and Rozen, A. (1995). Benchmarking Object-Oriented DBMSs for Workflow Management. *OOPSLA 95 Workshop on Object Database Behavior, Benchmarks and Performance*, Austin, USA.

Carey, M.J., Dewitt, D.J. and Naughton, J.F. (1993). The OO7 Benchmark. *ACM SIGMOD International Conference on Management of Data*, Washington, USA, 12-21.

Carey, M.J., Dewitt, D.J., Kant, C. and Naughton, J.F. (1994). A Status Report on the OO7 OODBMS Benchmarking Effort. *SIGPLAN Notices*, 29(10), 414-426.

Carey, M.J., Dewitt, D.J. and Naughton, J.F. (1997). The BUCKY Object-Relational Benchmark. *ACM SIGMOD International Conference on Management of Data*, Tucson, USA, 135-146.

Cattell, R.G.G. (1991). An Engineering Database Benchmark. *The Benchmark Handbook for Database Transaction Processing Systems*, Gray, J. (ed.): Morgan Kaufmann, 247-281.

Darmont, J., Petit, B. and Schneider, M. (1998). OCB: A Generic Benchmark to Evaluate the Performances of Object-Oriented Database Systems. *6$^{th}$ International Conference on Extending Database Technology*, Valencia, Spain. *LNCS*, 1377, 326-340.

Darmont, J. and Schneider, M. (2000). Benchmarking OODBs with a generic tool. *Journal of Database Management*, 11(3), 16-27.

Gerlhof, C., Kemper, A., Kilger, C. and Moerkotte, G. (1996). On the Cost of Monitoring and Reorganization of Object Bases for Clustering. *SIGMOD Record*, 25(3).

Gray, J. (ed.). (1993). "The Benchmark Handbook for Database and Transaction Processing Systems 2$^{nd}$ edition". Morgan Kaufmann.

He, Z. and Darmont, J. (2003). DOEF: A Dynamic Object Evaluation Framework. *14$^{th}$ International Conference on Database and Expert Systems Applications*, Prague, Czech Republic. *LNCS*, 2736, 662-671.

Kempe, J., Kowarschick, W., Kießling, W., Hitzelgerger, R. and Dutkowski, F. (1995). Benchmarking Object-Oriented Database Systems for CAD. *6$^{th}$*

*International Conference on Database and Expert Systems Applications*, London, UK. *LNCS*, 978, 167-176.

Kuno, H. and Rundensteiner, E.A. (1995). Benchmarks for Object-Oriented View Mechanisms. *OOPSLA 95 Workshop on Object Database Behavior, Benchmarks and Performance*, Austin, USA.

Lee, S., Kim, S. and Kim, W. (2000). The BORD Benchmark for Object-Relational Databases. *11<sup>th</sup> International Conference on Database and Expert Systems Applications*, London, UK. *LNCS*, 1873, 6-20.

Schreiber, H. (1994). JUSTITIA: a generic benchmark for the OODBMS selection. *4<sup>th</sup> International Conference on Data and Knowledge Systems in Manufacturing and Engineering*, Shatin, Hong Kong, 324-331.

Tiwary, A., Narasayya, V.R. and Levy, H.M. (1995). Evaluation of OO7 as a system and an application benchmark. *OOPSLA 95 Workshop on Object Database Behavior, Benchmarks and Performance*, Austin, USA.

Zimmermann, J. and Buchmann, A.P. (1995). Benchmarking Active Database Systems: A Requirement Analysis. *OOPSLA 95 Workshop on Object Database Behavior, Benchmarks and Performance*, Austin, USA.

## Terms and Definitions

**Object-oriented database:** A database system offering DBMS facilities in an object-oriented programming environment. Data are natively stored as objects.

**Object-relational database:** A database system where the relational model is extended with object-oriented concepts. Data are still stored in relational structures.

**Benchmark:** A standard program that runs on different systems to provide an accurate measure of their performance.

**Synthetic benchmark:** A benchmark in which the workload model is artificially generated, as opposed to a real-life workload.

**Database benchmark:** A benchmark specifically aimed at evaluating the performance of DBMSs or DBMS components.

**Workload model:** In a database benchmark, a database and a set of read and write operations to apply on this database.

**Performance metrics:** Simple or composite metrics aimed at expressing the performance of a system.